\documentclass[nopreprintline,11pt]{elsarticle}%

\usepackage[utf8]{inputenx}
\usepackage{indentfirst}
\usepackage{microtype}
\usepackage{caption}
\usepackage{subcaption}
\usepackage{setspace}
\usepackage{lettrine}
\usepackage{braket}
\usepackage[nottoc, notlof, notlot]{tocbibind}
\usepackage{hyperref}
\usepackage{amsmath}
\usepackage{amsfonts}
\usepackage{multicol}
\usepackage{wrapfig}
\usepackage{newpxtext,newpxmath}
\usepackage{setspace}
\usepackage[margin=1.25in]{geometry}
\usepackage{graphicx}
\usepackage{subcaption}
\usepackage{lineno}
\usepackage{xcolor}
\usepackage{placeins}
\usepackage{booktabs}
\usepackage{makecell}
\usepackage{tikz}
\usepackage{adjustbox}
\usepackage{quantikz}
\usepackage{natbib}
\biboptions{sort&compress}
\usepackage{bm}

\newcommand{\vect}[1]{\mathbf{#1}}
\newcommand{\gvect}[1]{\boldsymbol{#1}}

\begin{document}

\begin{frontmatter}

\title{Towards Efficient Quantum Hybrid Diffusion Models}
 
  \author[add1]{Francesca~De~Falco}
  \ead{francesca.defalco@uniroma1.it}
  \author[add1]{Andrea~Ceschini\corref{cor1}}
  \ead{andrea.ceschini@uniroma1.it}
  \author[add2]{Alessandro~Sebastianelli}
  \ead{alessandro.sebastianelli@esa.int}
  \author[add2]{Bertrand~Le~Saux}
  \ead{bertrand.le.saux@esa.int}
  \author[add1]{Massimo~Panella\corref{cor1}}
  \ead{massimo.panella@uniroma1.it}

  \cortext[cor1]{Please address correspondence to A. Ceschini or M. Panella}
  \address[add1]{Department of Information Engineering, Electronics and Telecommunications, University of Rome ``La Sapienza'', Italy}
  \address[add2]{ESA $\Phi$-Lab, European Space Agency, Frascati, Italy}

  \begin{abstract}
    In this paper, we propose a new methodology to design quantum hybrid diffusion models, derived from classical U-Nets with ResNet and Attention layers. Specifically, we propose two possible different hybridization schemes combining quantum computing's superior generalization with classical networks' modularity. In the first one, we acted at the vertex: ResNet convolutional layers are gradually replaced with variational circuits to create Quantum ResNet blocks. In the second proposed architecture, we extend the hybridization to the intermediate level of the encoder, due to its higher sensitivity in the feature extraction process. 
    In order to conduct an in-depth analysis of the potential advantages stemming from the integration of quantum layers, images generated by quantum hybrid diffusion models are compared to those generated by classical models, and evaluated in terms of several quantitative metrics. The results demonstrate an advantage in using a hybrid quantum diffusion models, as they generally synthesize better-quality images and converges faster. Moreover, they show the additional advantage of having a lower number of parameters to train compared to the classical one, with a reduction that depends on the extent to which the vertex is hybridized.
  \end{abstract}

  \begin{keyword}
    Quantum hybrid diffusion model, variational quantum circuit, quantum hybrid U-Net, efficient quantum simulation.
  \end{keyword}

\end{frontmatter}



\section{Introduction}
Quantum Machine Learning (QML) has recently emerged as a promising framework for generative Artificial Intelligence (AI). 
In fact, over the years, QML algorithms have been developed for both supervised \cite{farhi2018classification, johri:hal-03432449, Schuld_2017} and unsupervised \cite{aimeur:hal-00736948, Benedetti_2019} learning tasks.
The implementation and performance analysis of these machine learning algorithms have demonstrated that quantum computing can bring numerous advantages. In particular, the benefits come from the exponentially large space that a quantum system can express, as well as from the ability to represent mappings that are classically impossible to compute \cite{Bravyi_2018, Bravyi_2020}.
Additionally, in \cite{Abbas_2021} an investigation has been made whether quantum computing can bring benefits by achieving a better effective dimension than comparable classical neural networks.

In classical AI, Diffusion Models (DMs) have established themselves as the leading candidates for data and image generation \cite{pmlr-v37-sohl-dickstein15, ho2020denoising, Rombach2021HighResolutionIS}, showcasing superior quality and stability in training when compared to state-of-the-art Generative Adversarial Networks (GANs) \cite{NIPS2014_5ca3e9b1}. DMs rely on an iterative diffusion process that effectively models complex distributions by progressively refining the data distribution through a sequence of diffusion steps. However, DMs may encounter notable challenges, including high computational requirements and the need for extensive parameter adjustments \cite{dhariwal2021diffusion}.

Considering generative QML, there have been various implementations of Quantum Generative Adversarial Networks (QGANs) that have demonstrated superior performance in capturing the underlying data distribution. They have also shown better generalization properties, allowing for a significantly lower number of trainable parameters compared to classical GANs \cite{Huang_2021, latentquantumgans,Tsang2022HybridQG, NIPS2017_8a1d6947}.
One of the first implementations of a QGAN is described in \cite{Huang_2021}. In this implementation, only the generator is realized using quantum circuits, offering two different solutions: the first one, called Quantum Patch GAN, involves dividing an image into sub-images, which are then generated by sub-generators and later recombined to form the complete image, whereas in the second one, called Quantum Batch GAN, there is no longer a division into sub-images. 

Another possible implementation of QGANs is proposed in \cite{latentquantumgans}, where a novel architecture called Latent Style-based Quantum GAN (LaSt-QGAN) is introduced, i.e. a QGAN that operates directly in the latent space. 
In this scenario, the Generator is entirely quantum, with various implementations of the Generator circuit proposed, whereas the Discriminator remains entirely classical.
Authors of \cite{latentquantumgans} demonstrated that LaSt-QGAN outperforms a classical GAN with a comparable number of parameters in terms of metrics such as Fréchet Inception Distance (FID) and Jensen-Shannon Divergence (JSD). Furthermore, the quantum advantages become evident as the LaSt-QGAN achieves better metric values even in intermediate epochs and yields superior metric values when considering training on reduced percentages of the dataset.

Concerning DMs, there have been only some initial and simple attempts to develop a quantum version of a DM \cite{cacioppo2023quantum, parigi2023quantumnoisedriven}. While \cite{parigi2023quantumnoisedriven} solely offers a theoretical discussion of a potential quantum generalization of diffusion models, with results tied only to very basic and simplified scenarios, two different architectures are proposed in \cite{cacioppo2023quantum}: the first one works on downsized images from the MNIST dataset; the second model suggests operating on a latent space using a pretrained autoencoder. However, both architectures use amplitude encoding, which is highly inefficient, as it requires an exponential number of circuit runs to get the output distribution. Moreover, the simplicity of their architectures makes it challenging to extend them to more complex datasets, in contrast to the expressive richness of the U-Net commonly used in classical DMs.

In this paper, we propose an efficient methodology to design Hybrid Quantum Diffusion Models (HQDMs) by incorporating variational quantum layers with novel circuit designs within a classical U-Net \cite{10.1007/978-3-319-24574-4_28}. Specifically, we employ a state-of-the-art U-Net architecture, proposed in \cite{Rombach2021HighResolutionIS} and composed of ResNet \cite{He_2016_CVPR} and Attention blocks as the foundation for our approach. Therefore, the U-Net is hybridized in two different modes: the first involves inserting Variational Quantum Circuits (VQCs) only at the vertex, while the second implies the insertion of VQCs on both the encoder side, which is more sensitive to feature extraction, and the vertex, which is responsible for compressed image processing. The rationale behind this approach is to leverage the strengths of both classical and quantum computing paradigms. By integrating variational quantum layers into the classical U-Net architecture, we aim to exploit the expressive power of quantum circuits for faster network convergence and superior generalization capabilities \cite{Abbas_2021}. On the other hand, classical U-Net layers allow us to introduce modularity and nonlinearity in the computation, thus enabling complex image processing while mitigating the inherent challenges associated with quantum computing \cite{sebastianelli2021circuit,ceschini2022hybrid}. Furthermore, by strategically placing VQCs at key points within the U-Net architecture, we ensure that quantum layers are properly allocated to areas where they can have the most significant impact on performance improvement, thanks to their expressivity and feature extracting capabilities \cite{henderson2019quanvolutional,du2020expressive,wu2021expressivity}.

We also propose to adopt an approach inspired by transfer learning, aiming at an overall efficient training time of the models. During the initial epochs, a classical model is trained and then, some of its weights are transferred to a hybrid model that is trained, in turn, for some further epochs. In this way, while still maintaining a limited training time, we achieve better results compared to the classical counterpart: on Fashion MNIST, we achieve +2\% FID, +5\% KID, and +2\% IS. On MNIST, the results are even better with an improvement in metrics of approximately +8\% on FID, +11\% on KID, and +2\% on IS.

Our innovative architecture seamlessly incorporates quantum elements through the employment of angle encoding as the primary encoding method, whereas the outputs of the circuits are retrieved through the expected value of the Pauli-Z observable. This strategy is designed to achieve two crucial objectives: first, to keep the overall qubit count low, thereby optimizing resource utilization, and second, to streamline output computation without the need for an exponentially large number of circuit shots. This dual approach ensures not only the feasibility of implementation but also practicality when deploying our model on Noisy Intermediate-Scale Quantum (NISQ) devices.

Additionally, the analysis of two different hybrid architectures allows to highlight how certain parts of the U-Net are more sensitive to the quantum integration than others. 
Not only do these enhancements lead to remarkable improvements across various performance metrics, but they also yield a substantial reduction in the number of parameters requiring training. In particular, depending on the degree of U-Net hybridization, we can achieve up to about an 11\% reduction in parameters compared to the classical one. Furthermore, it is possible to achieve an improvement of almost 2\% on the FID and KID metrics in the case of the Fashion MNIST dataset, while in the case of the MNIST dataset, we achieve an improvement of about 5\% on the FID and more than 6\% on the KID compared to the classical network.This reduction not only streamlines the overall training process, but also contributes to more efficient utilization of computational resources, ultimately enhancing the model's scalability and applicability in real-world scenarios.

The rest of the paper is organized as follows. In Sect.~\ref{sec:background} we provide an explanation of DMs and variational circuits. In Sect.~\ref{sec: methodology} we present the employed methodology, while in Sect.~\ref{sec:results} we discuss the obtained results. Finally, we draw our conclusions in Sect.~\ref{sec: conclusions}.

\section{Theoretical Background}
\label{sec:background}
\subsection{Classical diffusion models}
In recent years, DMs have proven to be an important class of generative models. A standard mathematical formulation for diffusion models is the one presented by \cite{ho2020denoising} and here summarized to give to readers a general overview of its fundamentals. 

\begin{figure}[!ht]
    \centering
    \includegraphics[width=0.8\textwidth]{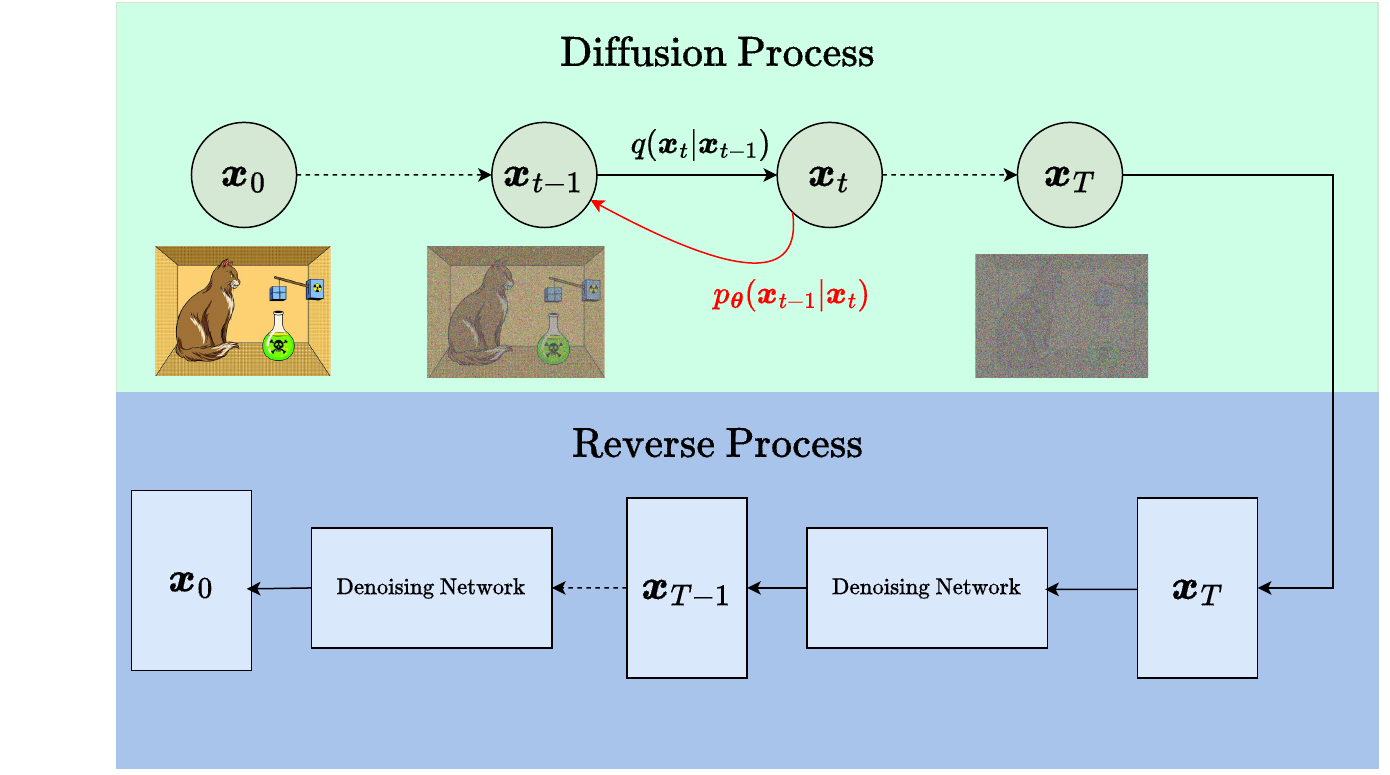}
    \caption{Diffusion process and reverse process of a DM.}
    \label{dm}
\end{figure}

DMs mainly consist of two distinct phases as shown in Fig. \ref{dm}. The first one is the forward process, also called diffusion, involving a transformation that gradually converts the original data distribution ${\vect{x}_0\sim q}$, where $q$ is a probability distribution to be learned, by repeatedly adding Gaussian noise:
\begin{equation}
	 \vect{x}_t = \sqrt{1 -\beta_{t}}\vect{x}_{t-1} + \sqrt{\beta_{t}}\gvect{\epsilon}_t,\,\,t=1\dots T\,,
\end{equation}
where ${\gvect{\epsilon}_1,\dots,\gvect{\epsilon}_T}$ are IID samples drawn from a zero-mean, unit variance Gaussian (normal) distribution $\mathcal{N}(\vect{0}, \vect{I})$, and $\beta_{t}$ determines the variance scale for the $t$-th step.
This progression is underpinned by a Markov chain that can be represented as follows:
\begin{align}
	&q(\vect{x}_{0:T}) = q(\vect{x}_0)\prod_{t=1}^{T}q(\vect{x}_{t}|\vect{x}_{t-1})\,,\\
	&q(\vect{x}_{t}|\vect{x}_{t-1})= \mathcal{N}\left(\sqrt{1 -\beta_{t}}\vect{x}_{t-1}, \beta_{t}\vect{I}\right)\,,	
\end{align}
being $\vect{I}$ the identity matrix. 

The goal of the forward process is to add incremental noise to the initial sample $\vect{x}_0$ over a certain number of steps, until at the final time step $T$ all traces of the original distribution ${\vect{x}_0\sim q}$ are lost so as to obtain ${\vect{x}_T\sim \mathcal{N}(\vect{0}, \vect{I})}$.
Through the application of the `reparameterization trick', a closed-form solution becomes available for calculating the total noise at any desired step using the cumulative product:
\begin{equation}
\vect{x}_{t} = \sqrt{\bar\alpha_{t}}\,\vect{x}_{0}+ \sqrt{1 -\bar\alpha_{t}}\,\gvect{\epsilon}\,,
\end{equation}
where $\alpha_{t} = 1 - \beta_{t}$, $\bar\alpha_{t}= \prod_{i=1}^{t} \alpha_{i}$, and $\gvect{\epsilon}\sim\mathcal{N}(\vect{0}, \vect{I})$ is the Gaussian noise.

The second phase is the reverse process or backward diffusion, where the transformations gradually restore the initial noise distribution and reconstruct a noise-free version of the original data. If we could successfully reverse the aforementioned process sampling from ${q(\vect{x}_{t-1}|\vect{x}_t)}$, we would gain the ability to recreate the true sample starting from the Gaussian noise input ${\vect{x}_T\sim\mathcal{N}(\vect{0}, \vect{I})}$; it is also noteworthy that when $\beta_t$ is sufficiently small, ${q(\vect{x}_{t-1}|\vect{x}_t)}$ is close to a Gaussian distribution. Regrettably, estimating ${q(\vect{x}_{t-1}|\vect{x}_t)}$ is complex due to its reliance on the entire dataset and hence, a data-driven learning model like a neural network must be used in order to approximate these conditional probabilities, enabling the execution of the reverse diffusion process.

Let $p_{\gvect{\theta}}$ be the mathematical model depending on some parameters $\gvect{\theta}$ that represents the estimated distribution of the backward diffusion process:
\begin{align}
	&p_{\gvect{\theta}}(\vect{x}_T) = \mathcal{N}(\vect{0}, \vect{I})\,,\\
	&p_{\gvect{\theta}}(\vect{x}_{t-1}|\vect{x}_{t})= \mathcal{N}\left(\gvect{\mu}_{\gvect{\theta}}(\vect{x}_{t},t), \gvect{\Sigma}_{\gvect{\theta}}(\vect{x}_{t},t)\right)\,,	
\end{align}
where $\gvect{\mu}_{\gvect{\theta}}(\vect{x}_{t},t)$ and $\gvect{\Sigma}_{\gvect{\theta}}(\vect{x}_{t},t)$ are the general outputs of the adopted neural network, which takes as inputs $\vect{x}_{t}$ and $t$.

A simplified approach based on variational inference assumes a fixed covariance matrix, such as for instance ${\gvect{\Sigma}_{\gvect{\theta}}(\vect{x}_{t},t)=\beta_t\vect{I}}$, and the direct estimation by the neural network of the noise $\gvect{\epsilon}_{\gvect{\theta}}(\vect{x}_{t},t)$ at time step $t$. Then, using reparameterization and the normal distributions of conditional data, we obtain:
\begin{equation}
	\gvect{\mu}_{\gvect{\theta}}(\vect{x}_{t},t) = {\frac{1}{\sqrt{\alpha_t}}} \left(\vect{x}_t - \frac{1-\alpha_t}{\sqrt{1-\bar\alpha_t}}\gvect{\epsilon}_{\gvect{\theta}}(\vect{x}_{t},t)\right)\,.
\end{equation}
The neural network producing $\gvect{\epsilon}_{\gvect{\theta}}(\vect{x}_{t},t)$ is usually trained by stochastic gradient descent on an even more simplified loss function like:
\begin{equation}
L^\mathrm{simple}_t = \mathbb{E}_{\vect{x}_0\sim q,t,\gvect{\epsilon}\sim\mathcal{N}(\vect{0},\vect{I})} \left[ \left\| \gvect{\epsilon}_{\gvect{\theta}}(\vect{x}_{t},t) - \gvect{\epsilon} \right\|^2 \right]\,.
\end{equation}
%
    

\subsection{Variational Quantum Circuits}
 Variational Quantum Algorithms (VQAs) are the most common QML algorithm that are currently implemented on today's quantum computers \cite{Cerezo_2021}, they make use of parametrized quantum circuits known as ansatzes. Ansatz circuits are composed of quantum gates that manipulate qubits through specific parametrized unitary operations. However, these operations depend on parameters denoted as $\gvect{\theta}$, which are the parameters to be trained during the training process.
 
\begin{figure}[!ht]
    \centering
    \includegraphics[trim={0 0 2.5cm 0}, width=0.8\textwidth]{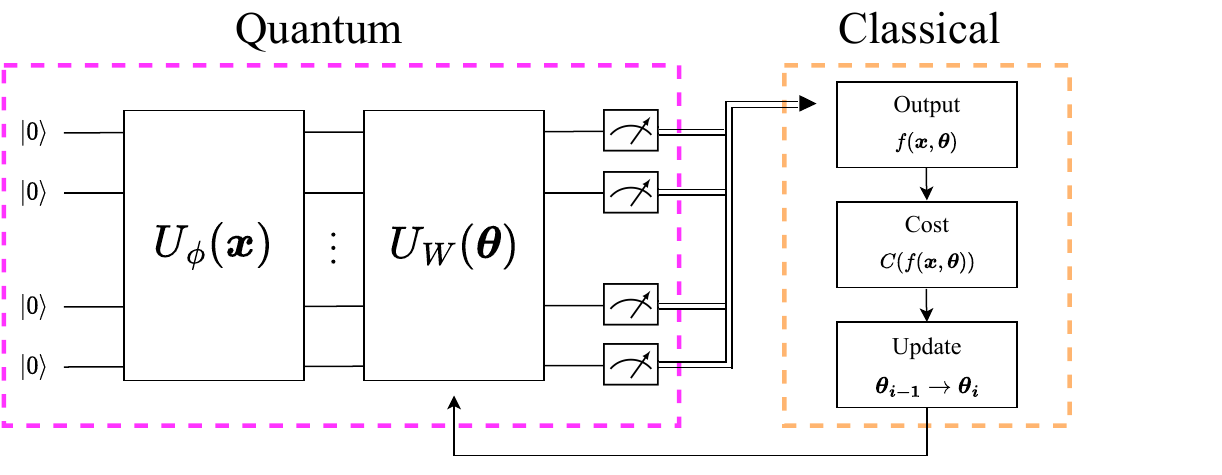}
    \caption{Scheme of a hybrid quantum-classical VQC.}
    \label{vqc}
\end{figure}

The training workflow of a VQC, shown in Fig.~\ref{vqc}, can be summarized as follows:
\begin{enumerate}
	\item Classical data $\gvect{x} \in \mathbb{R}^{n}$ is appropriately encoded in a quantum state of the Hilbert space $\mathbb{H}^{2^{n}}$ through the unitary $U_{\phi}({\gvect{x}})$, to be used by the quantum computer;
	\item An ansatz $U_W({\gvect{\theta}})$ of $\gvect{\theta}$-parametrized unitaries with randomly initialized parameters and fixed entangling gates is applied to the quantum state $|\phi({\gvect{x}})\rangle$ obtained after the encoding;
	\item Upon completion, measurements are taken to obtain the desired outcomes. The expected value with respect to a given observable $\hat{O}$ is typically computed, and the resulting prediction is given by: 
			\begin{equation}
			f({\gvect{x}}, \gvect{\theta}) = \langle \phi(\gvect{x}) | U_W(\gvect{\theta})^\dagger \hat{O} U_W(\gvect{\theta}) | \phi(\gvect{x}) \rangle\,,
			\end{equation}
    \item Finally, a suitable loss function is evaluated, and a classical co-processor is used to properly update the parameters $\gvect{\theta}$. 
\end{enumerate}
This cycle is repeated until a termination condition is met. To update $\gvect{\theta}$ and train the VQC, gradient-based techniques can be used; gradients in a parametrized quantum circuit are calculated via the parameter-shift rule:
\begin{equation}
     \nabla_{\theta}f(\gvect{x},\theta) = \frac{1}{2} \left[ f(\gvect{x},\theta + \frac{\pi}{2}) - f(\gvect{x},\theta - \frac{\pi}{2}) \right]\,,
\end{equation}
where $f(\gvect{x}, \theta)$ is the output of the quantum circuit and $\theta$ is the parameter to be optimized. 

\section{Proposed Methodology}
\label{sec: methodology}
In the following, we introduce the quantum hybrid architectures proposed in this paper.

\subsection{Quantum Vertex U-Net Hybrid Architecture}
\begin{figure}[!ht]
    \centering
    \includegraphics[width=0.95\textwidth]{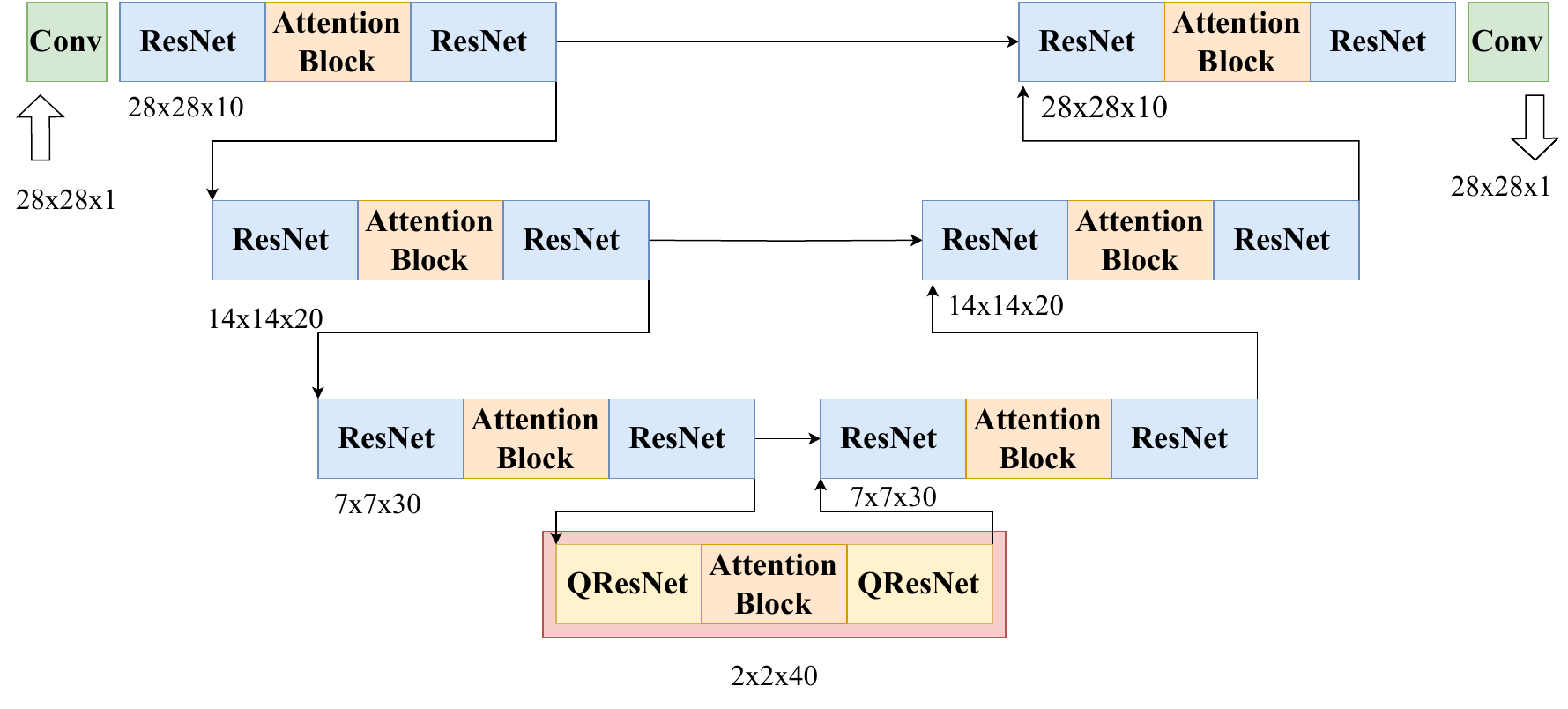}
    \caption{First proposed hybrid U-Net architecture named QVU-Net, where the quantum part is incorporated at the vertex of the network.}
    \label{firsthybrid}
\end{figure}

Starting from the classical U-Net architecture, which is an extremely parameter-rich and expressive model, we propose a hybrid architecture capable of harnessing quantum capabilities to achieve better performance and, furthermore, to reduce the total number of parameters to be trained.
The classical U-Net is initially implemented according to the proposal by \cite{pmlr-v139-nichol21a}, aiming at working with a state-of-the-art model: it is composed of ResNet blocks for residual connections and Attention blocks for feature aggregation; specifically, we use Multi-head Attention with four attention heads, as suggested in \cite{pmlr-v139-nichol21a}. The ResNet and Attention layers are hence applied at various resolution levels in the U-Net. 
The first hybrid architecture we propose, which we name Quantum Vertex U-Net (QVU-Net), uses this U-Net as its reference architecture and efficiently integrates quantum layers within its structure, as shown in Fig.~\ref{firsthybrid}. 
For this reason, we use angle encoding as our data encoding method, thus avoiding the use of amplitude encoding. Indeed, amplitude encoding is inefficient because it requires exponentially long circuits; it also necessitates an exponential number of circuit's run to generate statistically valid outputs from the quantum states distribution. In our angle encoding, input data $\gvect{x} \in \mathbb{R}^n$ is encoded into $n$ qubits via unitary transformation $U_{\phi(\gvect{x})}$ made up of $R_x$ rotation gates; each qubit encodes one feature of our input data.

The underlying idea in our hybrid model is to integrate the quantum components strategically, in such a way that the number of qubits used is kept limited while ensuring streamlined data processing efficiency. 
Specifically, considering that the classical U-Net initially processes a $28 \times 28 \times 1$ image and progressively scales it within the network's encoder until reaching a vertex with dimensions of $2 \times 2 \times 40$, we introduce the quantum elements precisely at the vertex of the network. 
%

\begin{figure}[!ht]
    \centering
    \includegraphics[width=1\linewidth]{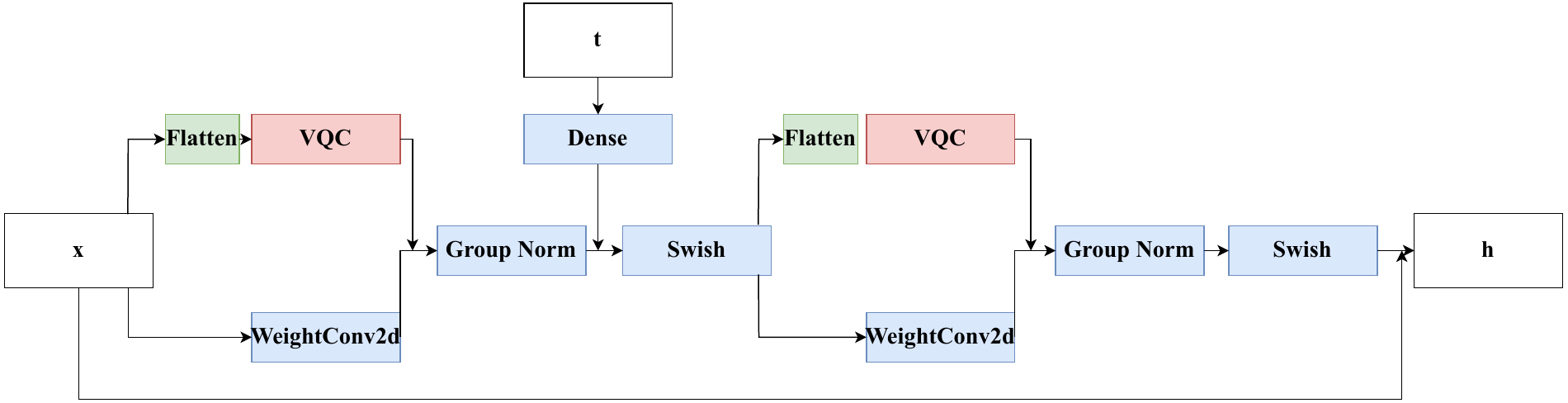}
    \caption{Architecture of the QResNet block, where Convolutional layers are replaced with VQCs. The QResNet takes as input x, which is the information coming from the image, and t, which is the temporal information, and finally returns h.}
    \label{qresnet}
\end{figure}

We propose the Quantum ResNet (QResNet) layer at the vertex of the QVU-Net, as shown in Fig.~\ref{qresnet}. The QResNet is analogous to the ResNet, as it is characterized by skip connections and two processing layers, whose output is then added to the input; the difference with the classical ResNet layer is that some of the Convolutional layers used in the classical ResNet are replaced with VQCs. As we are working with images scaled down to 2x2 dimensions, the only viable choice is to use a VQC instead of convolutional layers, which effectively amounts to a single filter pass over the entire image.
 Not all Convolutional layers of the ResNet are replaced with VQCs; rather, the replacement is done gradually. We initially analyze a hybrid architecture where the percentage of channels processed by VQCs in QResNet is set at 10\%. Subsequently, we examine an architecture in which 50\% of the input channels in QResNet are processed by VQCs. Finally, we explore an architecture in which all 40 channels are processed solely by VQCs instead of Convolutional layers. Gradually hybridizing the architecture allows us to better and more comprehensively analyze the impact of incorporating quantum elements into the network's structure. To the best of our knowledge, this is the first time a quantum version of the ResNet block within a hybrid U-Net is proposed in the literature.

The choice of the ansatz in the VQCs was made considering that the maximum channels at the vertex are 40. Since the information in the vertex is distributed across multiple channels rather than confined to one, we want to adopt an ansatz configuration capable of efficiently capturing and leveraging the correlations among these channels. Starting from these considerations, we try two distinct ansatzes structures, inspired by \cite{Jing_2022}. In fact, their choice is driven by the fact that they work on three channels simultaneously, aiming to process not only local information related to a single channel, but especially intra-channel information. 

\begin{figure}[!ht]
    \centering
    \includegraphics[trim={1cm, 0, 0.8cm, 0}, width=1. \textwidth]{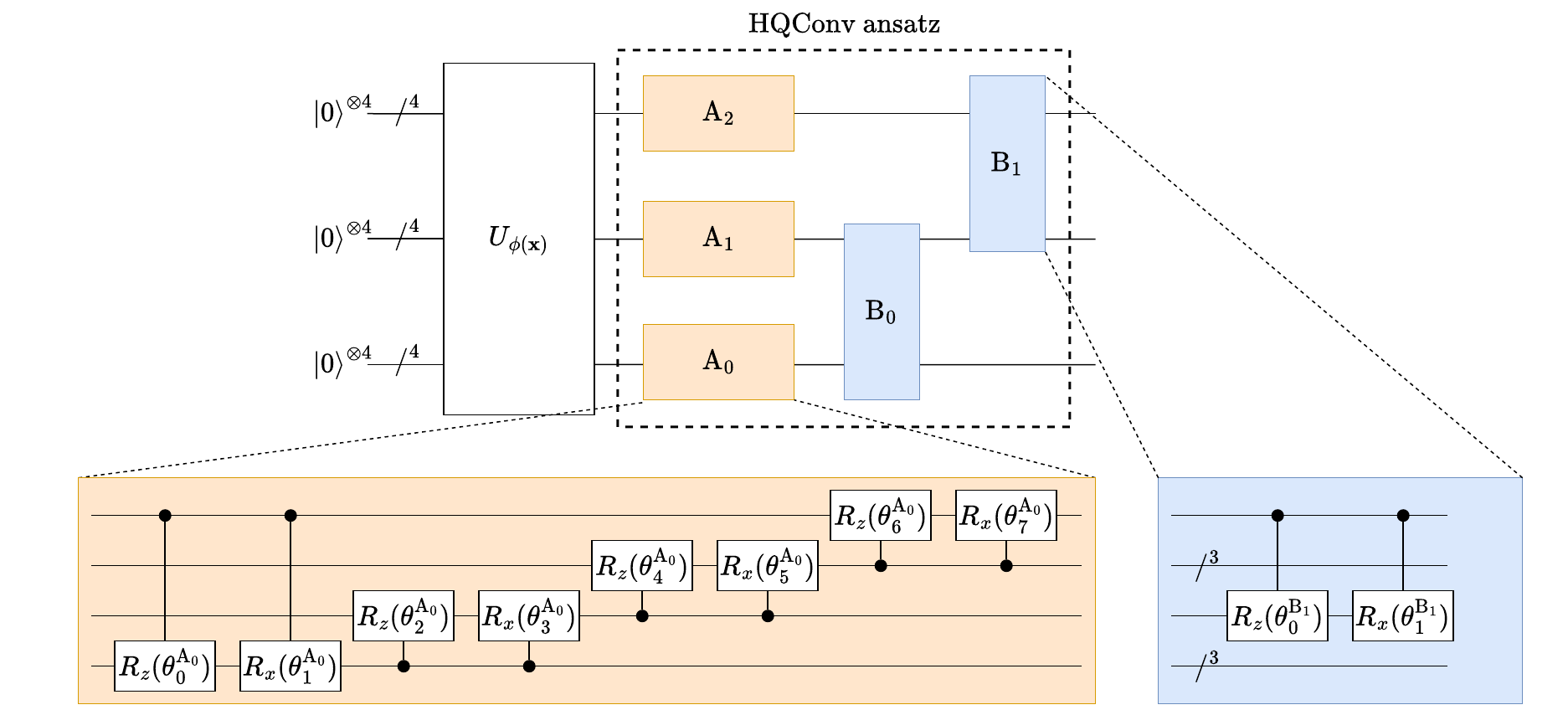}
    \caption{HQConv ansatz proposed in \cite{Jing_2022} initially extracts local information, then uses additional controlled gates to encode intra-channel information.}
    \label{hqconv}
\end{figure}

The Hierarchical Quantum Convolutional Ansatz (HQConv) in Fig.~\ref{hqconv} extracts local information separately among the initial channels, followed by additional controlled gates used to encode intra-channel information.
Initially, controlled gates are used to extract information within each channel first. In particular, as shown in Fig. \ref{hqconv}, the A blocks, acting on qubits belonging to the same channel, can be expressed mathematically as:
$$
\left|\, q^{2}_{p}, q^2_{ p+s}\right\rangle = \left(\, CR_x(\theta_{x,p}) \circ CR_z(\theta_{z,p})\,\right) \left|\, q^{1}_{p}, q^1_{ p+s}\right\rangle
$$
where $q^1_{p}$ is the control qubit and $q^1_{p+s}$ is the target qubit, moreover the subscript $p$ indicates the pixel to which the qubit refers and ranges from 0 to 3 as the images has dimensions of $2 \times 2$ for each channel. The symbol $s$ represents instead the value of hyperparameter stride, i.e. indicates the distance between the control qubit and the target qubit. The stride used in this case is equal to 1. The superscript 1 indicates the initial state of the qubit immediately after encoding, while the superscript 2 indicates the state of the qubit after the application of the considered block A.
The second part of the ansatz is instead characterized by intra-channel information processing. As seen in Fig. \ref{hqconv}, the B blocks are aimed at working on qubits belonging to two different channels, and in particular, we can express them in mathematical terms as:
$$
\left|\, q^{3}_{0}, q^3_{4}\right\rangle = \left(\, CR_x(\theta_{x}) \circ CR_z(\theta_{z})\,\right) \left|\, q^{2}_{0}, q^2_{4}\right\rangle
$$
where $q^2_0$ is the control qubit, i.e. the first qubit of the first channel considered in the block B and $q^2_4$ is the target qubit, i.e. the first qubit of the second channel considered in the block B. The subscript 3 in this case indicates the state of the qubit after the application of the block B.

\begin{figure}[!ht]
    \centering
    \includegraphics[trim={1.2cm, 0, 0.8cm, 0}, width=1.\textwidth]{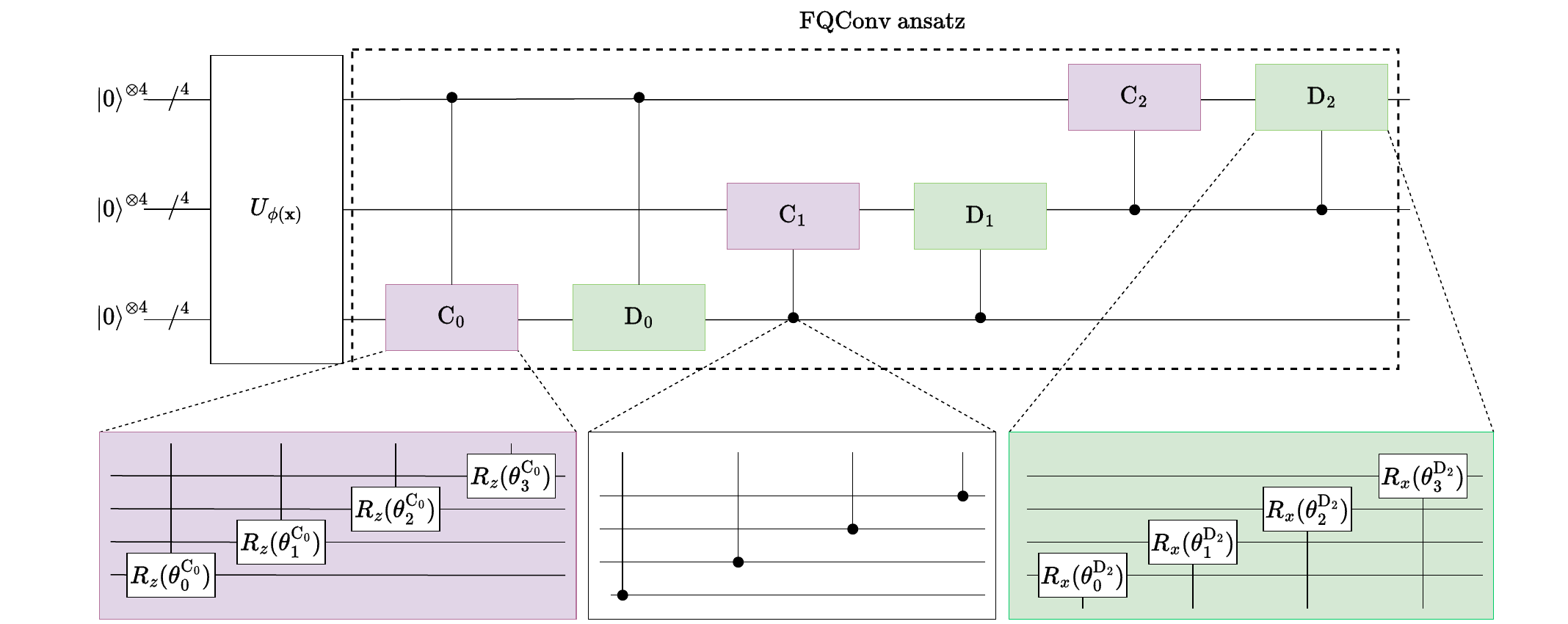}
    \caption{FQConv ansatz proposed in \cite{Jing_2022} is capable of immediately incorporating both intra-channel and inter-channel information.}
    \label{fqconv}
\end{figure}

On the other hand, the Flat Quantum Convolutional Ansatz (FQConv) in Fig.~\ref{fqconv} immediately incorporates both intra-channel and inter-channel information, giving its structure a flat form.
The blocks C and D, as shown in Fig. \ref{fqconv}, are characterized by the presence of gates controlled by qubits belonging to another channel. In mathematical terms, the block C can be expressed as:
$$
\left|\, q^{2}_{p}, q^2_{ p+s}\right\rangle = \left(\, CR_z(\theta_{z,p}) \right) \left|\, q^{1}_{p}, q^1_{ p+s}\right\rangle
$$
while the block D as:
$$
\left|\, q^{3}_{p}, q^3_{ p+s}\right\rangle = \left(\,  CR_x(\theta_{x,p})\,\right) \left|\, q^{2}_{p}, q^2_{ p+s}\right\rangle
$$
where as before $q_{p}$ is the control qubit and $q_{p+s}$ is the target qubit. The stride used in this case is equal to 4. Once again, the superscript 1 indicates the initial state of the qubit after encoding, 2 after the application of the first block (block C), and 3 the state of the qubit after the application of the second block (block D).
These ansatzes operate on 3 channels at a time; since the compressed image at the vertex is $2 \times 2$ in size, we need 4 qubits per channel, and hence the total number of qubits used in our VQCs is 12. The number of layer inside each VQC is equal to 3. At the end of the circuit, a measurement operation carried out in the Pauli-Z basis, which has +1 and -1 as eigenvalues, is performed on each qubit to extract the outcome of the quantum circuit. 

\subsection{Quanvolutional U-Net Hybrid Architecture}
\begin{figure}[!ht]
    \centering
    \includegraphics[width=0.95\linewidth]{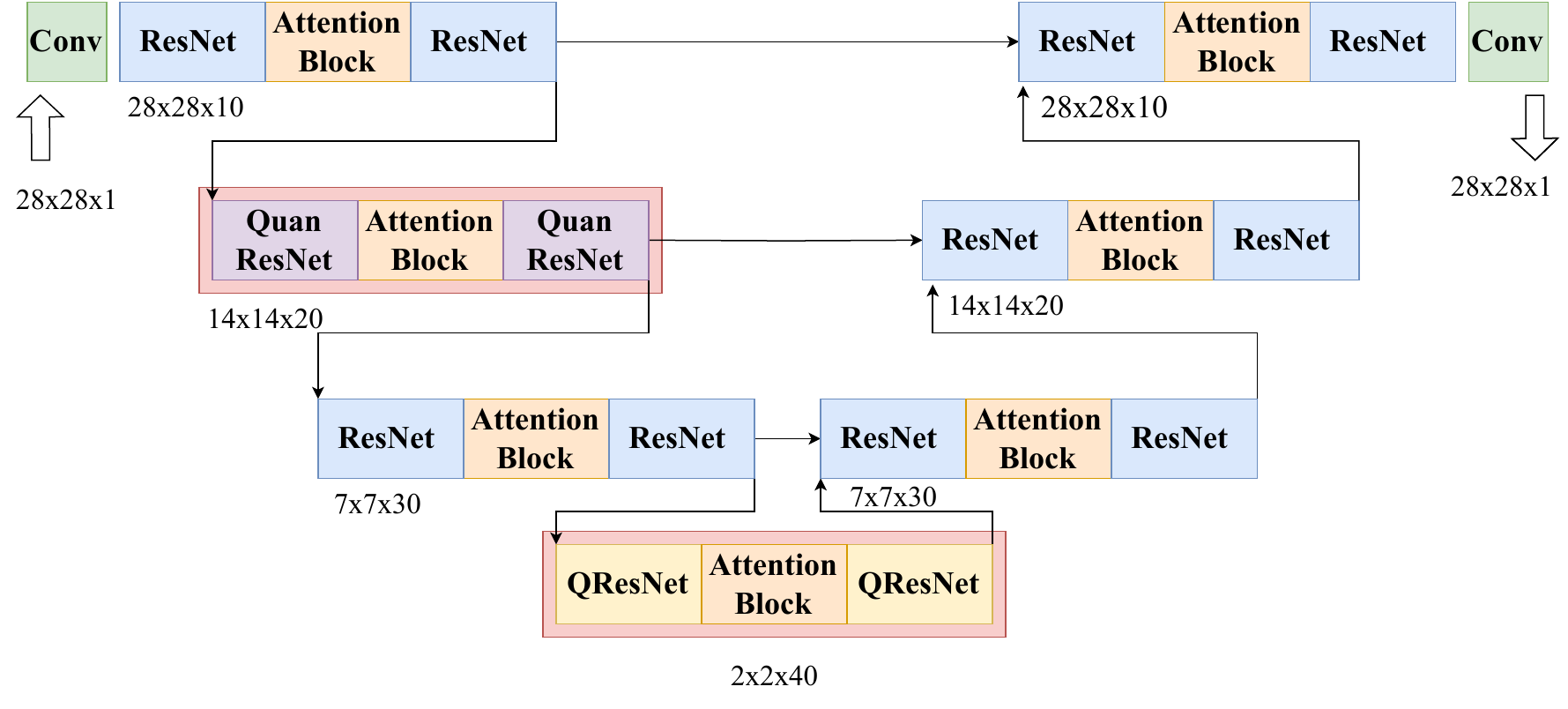}
    \caption{Second proposed U-Net hybrid architecture named QuanvU-Net, where the quantum part is incorporated not only at the vertex but also at the second level of the encoder block.}
    \label{secondhybrid}
\end{figure}

In addition to the QVU-Net, where only the vertex is hybridized, we also propose a hybridization in a part of the U-Net network dedicated to feature extraction. The purpose is to assess whether quantum feature extraction can indeed bring further improvements in terms of the quality of the generated images. However, we decide not to act on the first level of the U-Net because the images still have dimensions of $28 \times 28 \times 10$, making the efficient use of a VQC computationally challenging. Therefore, we consider the second level of the encoder, where the images are $14 \times 14 \times 20$, as shown in Fig.~\ref{secondhybrid}; we call this architecture QuanvU-Net.

\begin{figure}[!ht]
    \centering
    \includegraphics[width=1\linewidth]{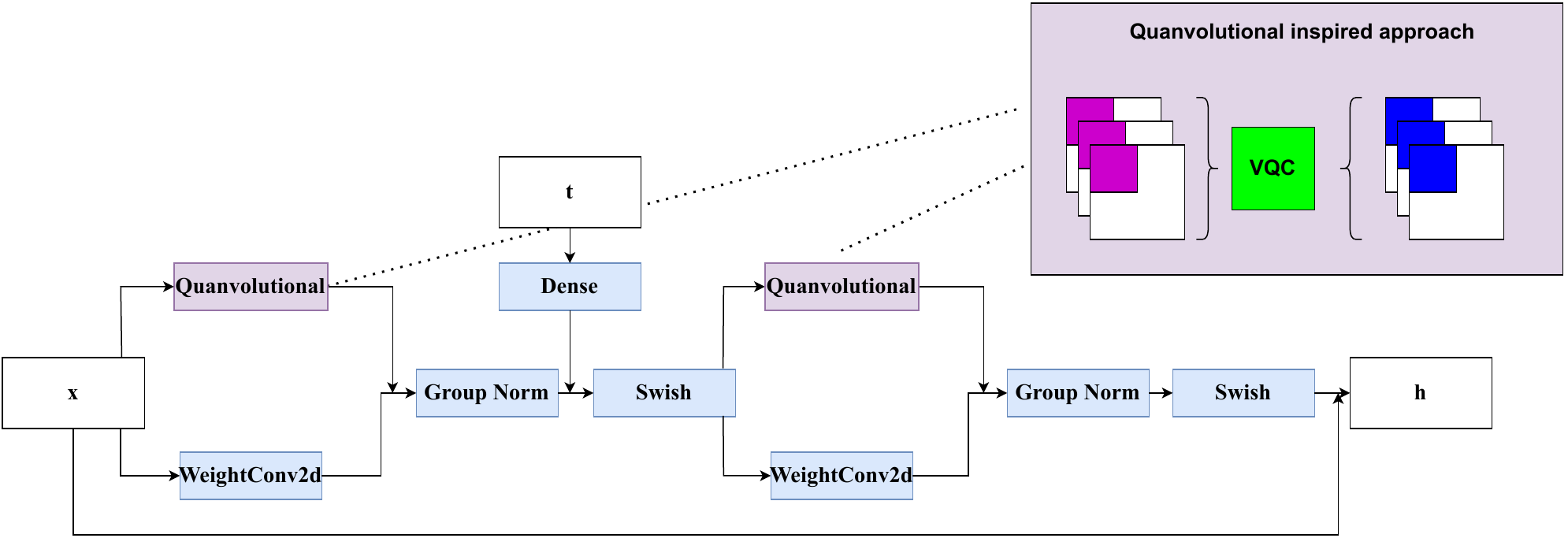}
    \caption{Architecture of the QuanResNet block, where the Convolutional layer of the classical ResNet is replaced with a Quanvolutional filter. The QuanResNet takes as input $\vect{x}$, which is the information coming from the image, and $\vect{t}$, which is the temporal information, and finally returns $\vect{h}$.}
    \label{quanresnet}
\end{figure}

In order to keep angle encoding, we process the image in the second level of the encoder with an idea inspired by the Quanvolutional method \cite{Henderson2019QuanvolutionalNN}. We employ Quanvolutional filters that, similarly to classical Convolutional filters, process one subsection of the image at a time, until they have traversed the entire image. In doing so, they produce a feature map by transforming spatially-local subsections of the input tensor.
Unlike the straightforward element-wise matrix multiplication operation performed by a classical Convolutional filter, a Quanvolutional filter alters input data through the utilization of a quantum circuit, which may have a structured or random configuration.
In our case, the Quanvolutional approach is employed within the ResNet block, thereby creating the QuanResNet block, as depicted in Fig.~\ref{quanresnet}.
Specifically, we consider only 3 channels out of the total 20 in the $14 \times 14$ image to make the approach practically feasible, as it requires a large number of quantum circuits executions. In particular, 4 pixels are taken from each channel and processed by a 12-qubits variational circuit; the variational circuit remains always the same when passing over the entire image and acts as if it was a classical Convolutional filter. 

As for the previous QVU-Net architecture, the idea with the QuanvU-Net is to work on 3 channels at a time to process both intra-channel and inter-channel features. Therefore, the ansatzes used in the QuanResNet block are the same as the ones used at the vertex, namely HQConv and FQConv shown in Fig.~\ref{hqconv} and Fig.~\ref{fqconv}, respectively. Also in this case, the use of Quanvolutional filters in a ResNet block marks a novel advancement compared to the literature.

\subsection{Transfer Learning Approach}
In addition to standard training for both the hybrid and classical networks, we propose an approach inspired by transfer learning. In fact, both the training phase and the subsequent inference phase prove to be significantly time consuming in the case of the hybrid QVU-Net and QuanvU-Net, becoming even longer compared to the classic U-Net as the percentage of variational circuits at the vertex increases. It is therefore worth reducing the time required for the training phase.

\begin{figure}[!ht]
    \centering
    \includegraphics[width=0.85\textwidth]{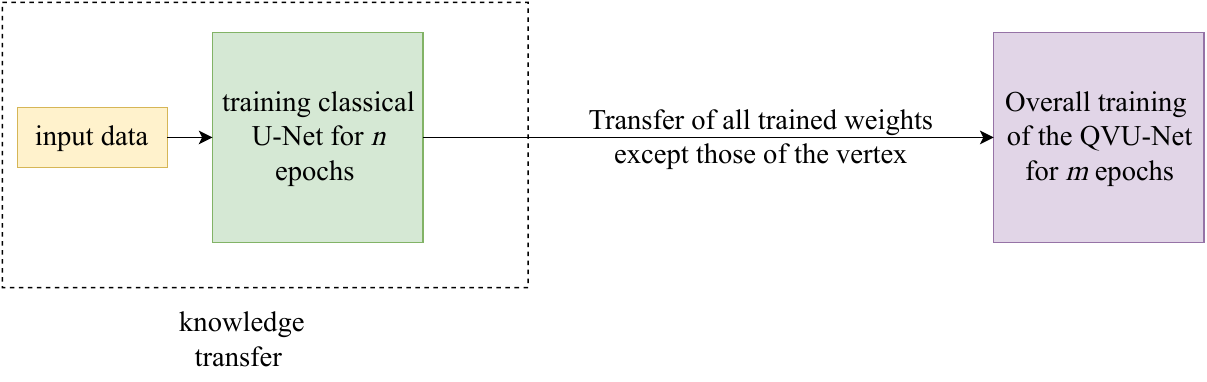}
    \caption{Outline of the proposed transfer learning approach. The classical model is trained for $n$ epochs. All the weights, except those of the vertex, are then transferred to the hybrid model. The latter is further trained for additional $m$ epochs.}
    \label{transferlearning}
\end{figure}

For this reason, we propose to adapt classical-to-quantum transfer learning \cite{Mari_2020}, which proves to be one of the most appealing transfer learning approaches. 
As illustrated in Fig.~\ref{transferlearning}, our idea is to initially train the classic U-Net for a certain number of epochs. The weights obtained in this way are then transferred to the QVU-Net, except for the weights at the vertex. So the QVU-Net is trained for a very limited number of epochs. In this way, it is expected that the final fine-tuning with the hybrid networks can bring a significant improvement in performance compared to using only the classical network, while still maintaining a low training time.



\section{Experimental Results}
\label{sec:results}
In this section, we analyze the results obtained by using the different architectures proposed in the paper: 
\begin{itemize}
\item the 1HQConv QVU-Net and 1FQConv QVU-Net with quantum circuits in the QResNet that act only on 10\% of the channels at the vertex and use the HQConv or FQConv the ansatz, respectively;
\item  7HQConv QVU-Net and 7FQConv QVU-Net, with quantum circuits in the QResNet that act only on 50\% of the channels at the vertex and use the HQConv or FQConv the ansatz, respectively;
\item FullHQConv QVU-Net and FullFQConv QVU-Net, with quantum circuits in the QResNet that act on all channels at the vertex and uses the HQConv or FQConv the ansatz, respectively;
\item QuanvU-Net with the vertex ibridized at 10\% and the QuanResNet at the second level of the encoder.
\end{itemize}

\subsection{Experimental settings}
The implementation is carried out in Python 3.8 using PennyLane and Flax. PennyLane is a framework which enables local quantum circuits simulations and integration with classical neural networks, whereas Flax is an open-source machine learning framework that provides a flexible and efficient platform for hybrid neural network execution via compilation. 
We use PennyLane{\textsuperscript\texttrademark} for the implementation of quantum circuits, while the hybrid networks and the entire training process are carried out in Flax; the classical U-Net is implemented in Flax as well.

Regarding the experiments, the L2 loss is used with the P2 weighting \cite{choi2022perception}. We used an exponential moving average (EMA) over model parameters with a rate that depends on the training step, and the Adam optimizer \cite{kingma2014adam} is used with a learning rate of $10^{-3}$, $\beta_1$ of 0.9, and $\beta_2$ of 0.99. The training process consists of a total of 20 epochs.

We use two benchmark datasets, namely MNIST \cite{lecun2010mnist} and Fashion MNIST \cite{DBLP:journals/corr/abs-1708-07747}. Both of them contain grayscale images belonging to 10 different classes, with a total of 60k training samples.
The metrics used for evaluations are FID \cite{NIPS2017_8a1d6947}, Kernel Inception Distance (KID) \cite{NIPS2016_8a3363ab}, and Inception Score (IS) \cite{betzalel2022study}, assessed on 7000 generated images. We utilize the TorchMetrics library \cite{Detlefsen2022}, replicating each channel of the generated images three times to make the dimensions compatible with those required by InceptionV3 network backbone.

A machine equipped with an AMD Ryzen {7\textsuperscript\texttrademark} 5800X 8-Core CPU at 3.80 GHz and with 64 GB of RAM is used for the experiments.

\subsection{Fashion MNIST dataset}
\begin{figure}[!ht]
  \centering
  \begin{subfigure}{0.23\textwidth}
    \includegraphics[width=\linewidth]{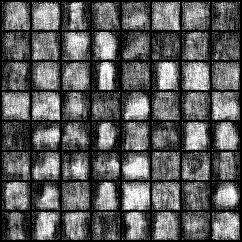}
    \caption{\scriptsize
    Classical  \\
    Params= 483321\\
    FID= 296.869\\
    KID= 0.3656 $\pm$ 0.0024\\
    IS= 1.4164 $\pm$ 0.0122 }
    \label{fig:classica_fashion_1}
  \end{subfigure}
  \hfill
  \begin{subfigure}{0.23\textwidth}
    \includegraphics[width=\linewidth]{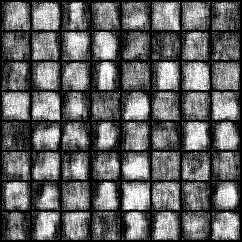}
    \caption{ \scriptsize
    1HQConv QVU-Net  \\
    Params= 475329 (-1.66\%)\\
    FID= 295.9013\\
    KID= 0.3646 $\pm$ 0.0023\\
    IS= 1.4389 $\pm$ 0.0119}
    \label{fig:1hqconv_3_fashion_1}
  \end{subfigure}
  \hfill
  \begin{subfigure}{0.23\textwidth}
    \includegraphics[width=\linewidth]{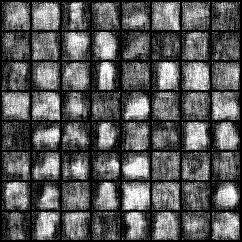}
    \caption{ \scriptsize
    7HQConv QVU-Net  \\
    Params= 440985 (-8.76\%)\\
    FID= 295.7298\\
    KID= 0.3647 $\pm$ 0.0026\\
    IS= 1.4273 $\pm$ 0.0142}
    \label{fig:7hqconv_3_fashion_1}
  \end{subfigure}
 \hfill
  \begin{subfigure}{0.23\textwidth}
    \includegraphics[width=\linewidth]{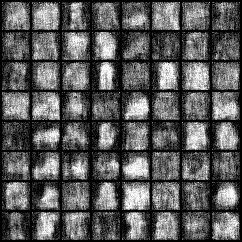}
    \caption{\scriptsize FullHQConv QVU-Net  \\
    Params= 429993 (-11.03\%)\\
    FID= 295.7863\\
    KID= 0.3692 $\pm$ 0.0025\\
    IS= 1.4454 $\pm$ 0.0057 }
    \label{fig:fullhqconv_3_fashion_1}
  \end{subfigure}
   \hfill
  \begin{subfigure}{0.23\textwidth}
    \includegraphics[width=\linewidth]{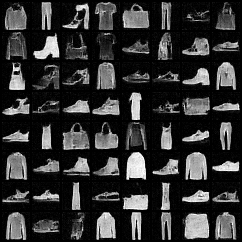}
    \caption{ \scriptsize
    Classical  \\
    Params= 483321\\
    FID= 60.1476\\
    KID= 0.0489 $\pm$ 0.0013\\
    IS= 3.8841$\pm$ 0.1174}
    \label{fig:classica_fashion_10}
  \end{subfigure}
   \hfill
  \begin{subfigure}{0.23\textwidth}
    \includegraphics[width=\linewidth]{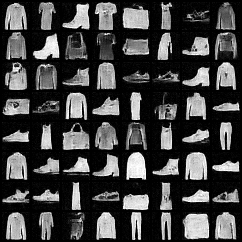}
    \caption{\scriptsize
    1HQConv QVU-Net  \\
    Params= 475329 (-1.66\%)\\
    FID= 52.5332\\
    KID= 0.0411$\pm$ 0.0010\\
    IS=  3.8667$\pm$ 0.1065 }
    \label{fig:1hqconv_3_fashion_10}
  \end{subfigure}
   \hfill
  \begin{subfigure}{0.23\textwidth}
    \includegraphics[width=\linewidth]{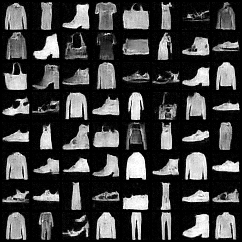}
    \caption{ \scriptsize
    7HQConv QVU-Net  \\
    Params= 440985 (-8.76\%)\\
    FID= 53.9605\\
    KID= 0.0420 $\pm$ 0.0011\\
    IS= 3.7905 $\pm$ 0.1323}
    \label{fig:7hqconv_3_fashion_10}
  \end{subfigure}
   \hfill
  \begin{subfigure}{0.23\textwidth}
    \includegraphics[width=\linewidth]{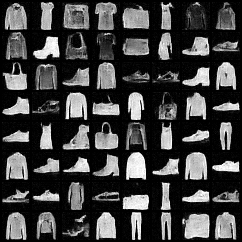}
    \caption{\scriptsize FullHQConv QVU-Net  \\
    Params= 429993 (-11.03\%)\\
    FID= 56.4960\\
    KID= 0.0433 $\pm$ 0.0010 \\
    IS= 3.8691 $\pm$ 0.0580}
    \label{fig:fhqconv_3_fashion_10}
  \end{subfigure}
   \hfill
  \begin{subfigure}{0.23\textwidth}
    \includegraphics[width=\linewidth]{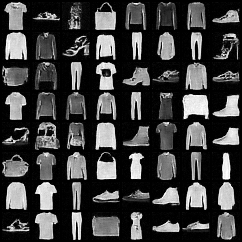}
    \caption{\scriptsize
    Classical  \\
    Params= 483321\\
    FID= 39.4563\\
    KID= 0.0275 $\pm$ 0.0008\\
    IS= 3.9783 $\pm$ 0.0777 }
    \label{fig:classica_fashion_20}
  \end{subfigure}
   \hfill
  \begin{subfigure}{0.23\textwidth}
    \includegraphics[width=\linewidth]{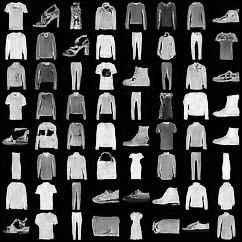}
    \caption{ \scriptsize 
    1HQConv QVU-Net  \\
    Params= 475329 (-1.66\%)\\
    FID= 39.9935\\
    KID= 0.0278 $\pm$ 0.0008\\
    IS= 3.9787 $\pm$ 0.0550}
    \label{fig:1hqconv_fashion_20}
  \end{subfigure}
   \hfill
    \hfill
  \begin{subfigure}{0.23\textwidth}
    \includegraphics[width=\linewidth]{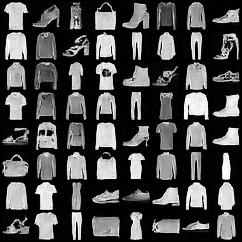}
    \caption{\scriptsize 7HQConv QVU-Net  \\
    Params= 440985 (-8.76\%)\\
    FID= 40.3685\\
    KID= 0.0281 $\pm$ 0.0009\\
    IS= 3.8158 $\pm$ 0.1360 }
    \label{fig:7hqconv_fashion_20}
  \end{subfigure}
  \hfill
  \begin{subfigure}{0.23\textwidth}
    \includegraphics[width=\linewidth]{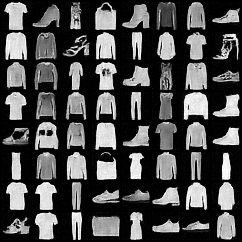}
    \caption{ \scriptsize FullHQConv QVU-Net  \\
    Params= 429993 (-11.03\%)\\
    FID= 41.3882\\
    KID= 0.0294 $\pm$ 0.0009\\
    IS= 3.8019 $\pm$ 0.0744}
    \label{fig:fhqconv_fashion_20}
  \end{subfigure}
  \caption{Fashion MNIST dataset results using the first hybrid architecture. The figure shows in the first column the images generated by the classical U-Net network, while in the second column the images generated by the 1HQConv QVU-Net , in the third column by the 7HQConv QVU-Net, and in the last column the images generated by the FullHQConv QVU-Net. The row-wise division considers in the first row the images generated after the networks are trained for just one epoch, the second row after training for ten epochs, and the third row after the complete training of twenty epochs. 
}
  \label{fig:fashion_hqconv_3}
\end{figure}

Let's initially consider the images generated by Fashion MNIST. As there are no significant differences between the use of HQConv and FQConv, we specifically examine the results obtained by HQConv with three layers. As seen in Fig.~\ref{fig:fashion_hqconv_3}, at the first epoch hybrid networks demonstrate better performance compared to the classical one, achieving a slightly lower FID and a higher IS value. This is in line with our expectations, as quantum models with a few epochs are generally more adept at extracting features and processing than classical models. By the tenth epoch, all hybrid networks show significantly better values in terms of FID and KID, with the 1HQConv QVU-Net Architecture in Fig.~\ref{fig:1hqconv_3_fashion_10} having an FID more than seven points lower than the classical network in Fig.~\ref{fig:classica_fashion_10}. The IS of hybrid networks at the tenth epoch is comparable to that achieved by the classical network. However, at the twentieth epoch, a gradual deterioration in the performance of hybrid networks in Figs.~\ref{fig:1hqconv_fashion_20}, \ref{fig:7hqconv_fashion_20}, \ref{fig:fhqconv_fashion_20} is observed, progressively worsening with an increase in the level of hybridization. Nevertheless, the results are still comparable to those obtained by the classical network and, most importantly, there is a significant reduction in parameters as the percentage of vertex hybridization increases.

\begin{figure}[!ht]
  \centering
  \begin{subfigure}{0.3\textwidth}
    \includegraphics[width=\linewidth]{classica_l2_1.png}
    \caption{\scriptsize
    Classical  \\
    Params= 483321\\
    FID= 296.869\\
    KID= 0.3656 $\pm$ 0.0024\\
    IS= 1.4164 $\pm$ 0.0122 }
    \label{fig:classica_fashion_1_2}
  \end{subfigure}
  \hfill
  \begin{subfigure}{0.3\textwidth}
    \includegraphics[width=\linewidth]{4ansatz1_3_1.png}
    \caption{ \scriptsize
    1HQConv QVU-Net  \\
    Params= 475329 (-1.66\%)\\
    FID= 295.9013\\
    KID= 0.3646 $\pm$ 0.0023\\
    IS= 1.4389 $\pm$ 0.0119}
    \label{fig:1hqconv_3_fashion_1_2}
  \end{subfigure}
  \hfill
  \begin{subfigure}{0.3\textwidth}
    \includegraphics[width=\linewidth]{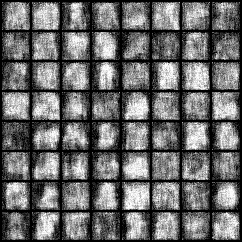}
    \caption{ \scriptsize
    QuanvU-Net \\
    Params= 474293 (-1.86\%) \\
    FID= 285.1551\\
    KID= 0.3521 $\pm$ 0.0024\\
    IS= 1.4309 $\pm$ 0.0148}
    \label{fig:quan_fashion_1}
  \end{subfigure}
 \hfill
  \begin{subfigure}{0.3\textwidth}
    \includegraphics[width=\linewidth]{classica_l2_10.png}
    \caption{ \scriptsize
    Classical  \\
    Params= 483321\\
    FID= 60.1476\\
    KID= 0.0489 $\pm$ 0.0013\\
    IS= 3.8841$\pm$ 0.1174}
    \label{fig:classica_fashion_10_2}
  \end{subfigure}
   \hfill
  \begin{subfigure}{0.3\textwidth}
    \includegraphics[width=\linewidth]{4ansatz1_3_10.png}
    \caption{\scriptsize
    1HQConv QVU-Net  \\
    Params= 475329 (-1.66\%)\\
    FID= 52.5332\\
    KID= 0.0411$\pm$ 0.0010\\
    IS=  3.8667$\pm$ 0.1065 }
    \label{fig:1hqconv_3_fashion_10_2}
  \end{subfigure}
   \hfill
  \begin{subfigure}{0.3\textwidth}
    \includegraphics[width=\linewidth]{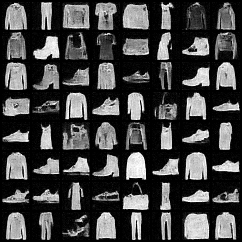}
    \caption{ \scriptsize
    QuanvU-Net \\
    Params= 474293 (-1.86\%)\\
    FID= 53.9734\\
    KID= 0.0427 $\pm$ 0.001\\
    IS= 3.7784 $\pm$ 0.0563}
    \label{fig:quan_fashion_10}
  \end{subfigure}
   \hfill
  \begin{subfigure}{0.3\textwidth}
    \includegraphics[width=\linewidth]{classica_l2_20.png}
    \caption{\scriptsize
    Classical  \\
    Params= 483321\\
    FID= 39.4563\\
    KID= 0.0275 $\pm$ 0.0008\\
    IS= 3.9783 $\pm$ 0.0777 }
    \label{fig:classica_fashion_20_2}
  \end{subfigure}
   \hfill
  \begin{subfigure}{0.3\textwidth}
    \includegraphics[width=\linewidth]{4ansatz1_3_20.png}
    \caption{ \scriptsize 
    1HQConv QVU-Net  \\
    Params= 475329 (-1.66\%)\\
    FID= 39.9935\\
    KID= 0.0278 $\pm$ 0.0008\\
    IS= 3.9787 $\pm$ 0.0550}
    \label{fig:1hqconv_fashion_20_2}
  \end{subfigure}
   \hfill
    \hfill
  \begin{subfigure}{0.3\textwidth}
    \includegraphics[width=\linewidth]{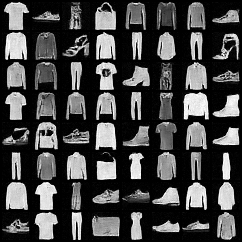}
    \caption{\scriptsize QuanvU-Net  \\
    Params= 474293 (-1.86\%) \\
    FID= 38.8\\
    KID= 0.0269$\pm$ 0.0007\\
    IS= 3.9087 $\pm$ 0.1485}
    \label{fig:quan_fashion_20}
  \end{subfigure}
  \caption{Fashion MNIST dataset results using the second hybrid architecture. The figure shows in the first column the images generated by the classical U-Net network, while in the second column the images generated by the 1HQConv QVU-Net, in the third column by the QuanvU-Net. The row-wise division considers in the first row the images generated after the networks are trained for just one epoch, the second row after training for ten epochs, and the third row after the complete training of twenty epochs.}
  \label{fig:fashion_hqconv_3_2}
\end{figure}

Considering the images generated by the second possible hybridization of the U-Net, the QuanvU-Net architecture, which involves not only the hybridized vertex but also the use of quanvolutional on outer layers of the U-Net, it can be observed in Fig.~\ref{fig:fashion_hqconv_3_2} how this yields a better performance. Considering the case of only 10\% of the hybridized vertex, which leads to more satisfactory performance, we outline that if only the vertex is hybridized at the first epoch, the metrics are more or less similar to those of the classical network. However, if we also consider the QuanvU-Net from the first epoch, as shown in Fig.~\ref{fig:quan_fashion_1}, the performance is greatly improved with a FID lower than that of the classical network by about 10 points. By the tenth epoch, there are no significant differences between the two possible implementations. More interesting is the case of the last epoch Fig.~\ref{fig:quan_fashion_20}, where initially the introduction of the quantum did not bring any improvement. Now, however, better performance in terms of FID and KID is achieved. This is in line with our expectations, given that we have now incorporated a quantum component into a much more critical area of the U-Net, which not only processes but, more importantly, extracts features.

\subsection{MNIST dataset}
\begin{figure}[!ht]
  \centering
  \begin{subfigure}{0.23\textwidth}
    \includegraphics[width=\linewidth]{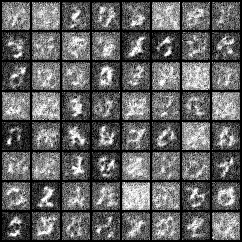}
    \caption{\scriptsize 
    Classical  \\
    Params= 483321\\
    FID= 311.0582\\
    KID= 0.4144 $\pm$ 0.0029\\
    IS= 1.4936 $\pm$ 0.0142 }
    \label{fig:mnist_classica_1}
  \end{subfigure}
  \hfill
  \begin{subfigure}{0.23\textwidth}
    \includegraphics[width=\linewidth]{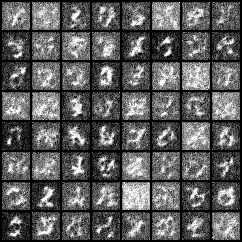}
    \caption{ \scriptsize
    1HQConv QVU-Net  \\
    Params= 475329 (-1.66\%)\\
    FID= 305.7763\\
    KID= 0.4087 $\pm$ 0.0033\\
    IS= 1.5264 $\pm$ 0.0185}
    \label{fig:mnist_1_1}
  \end{subfigure}
  \hfill
  \begin{subfigure}{0.23\textwidth}
    \includegraphics[width=\linewidth]{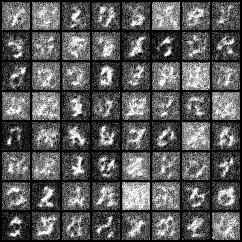}
    \caption{ \scriptsize 7HQConv QVU-Net  \\
    Params= 440985 (-8.76\%)\\
    FID= 302.5418\\
    KID= 0.4024 $\pm$ 0.0030\\
    IS= 1.5697 $\pm$ 0.0267}
    \label{fig:mnist_7_1}
  \end{subfigure}
 \hfill
  \begin{subfigure}{0.23\textwidth}
    \includegraphics[width=\linewidth]{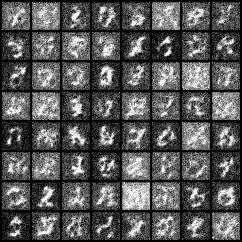}
    \caption{\scriptsize 
    FullHQConv QVU-Net  \\
    Params= 429993 (-11.03\%)\\
    FID= 299.7292\\
    KID= 0.3974 $\pm$ 0.0028\\
    IS= 1.5930 $\pm$ 0.0239 }
    \label{fig:mnist_f_1}
  \end{subfigure}
   \hfill
  \begin{subfigure}{0.23\textwidth}
    \includegraphics[width=\linewidth]{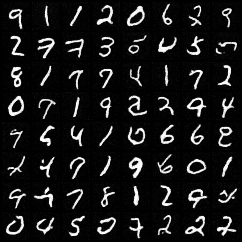}
    \caption{ \scriptsize Classical  \\
    Params= 483321\\
    FID= 75.7591\\
    KID= 0.098$\pm$ 0.0022\\
    IS= 1.8097 $\pm$ 0.0205}
    \label{fig:mnist_classica_10}
  \end{subfigure}
   \hfill
  \begin{subfigure}{0.23\textwidth}
    \includegraphics[width=\linewidth]{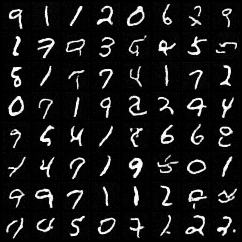}
    \caption{\scriptsize 1HQConv QVU-Net  \\
    Params= 475329 (-1.66\%)\\
    FID= 78.6997\\
    KID= 0.1011 $\pm$ 0.0024\\
    IS= 1.7940 $\pm$ 0.0345}
    \label{fig:mnist_1_10}
  \end{subfigure}
   \hfill
  \begin{subfigure}{0.23\textwidth}
    \includegraphics[width=\linewidth]{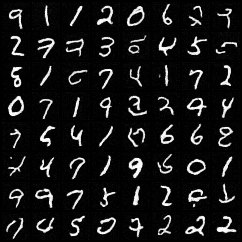}
    \caption{ \scriptsize 7HQConv QVU-Net  \\
    Params= 440985 (-8.76\%)\\
    FID= 74.6193\\
    KID= 0.0960 $\pm$ 0.0022\\
    IS= 1.8197 $\pm$ 0.0275}
    \label{fig:mnist_7_10}
  \end{subfigure}
   \hfill
  \begin{subfigure}{0.23\textwidth}
    \includegraphics[width=\linewidth]{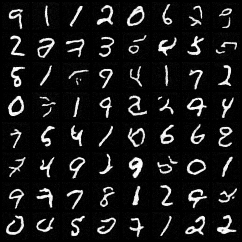}
    \caption{\scriptsize FullHQConv QVU-Net  \\
    Params= 429993 (-11.03\%)\\
    FID= 78.3131\\
    KID= 0.1036 $\pm$ 0.0023\\
    IS= 1.8008$\pm$ 0.0203}
    \label{fig:mnist_f_10}
  \end{subfigure}
   \hfill
  \begin{subfigure}{0.23\textwidth}
    \includegraphics[width=\linewidth]{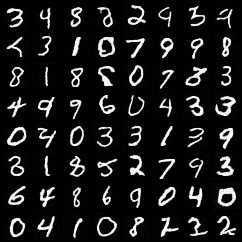}
    \caption{\scriptsize Classical  \\
    Params= 483321\\
    FID= 45.2762\\
    KID= 0.0558 $\pm$ 0.0014\\
    IS= 1.9152 $\pm$ 0.0355 }
    \label{fig:mnist_classica_20}
  \end{subfigure}
   \hfill
  \begin{subfigure}{0.23\textwidth}
    \includegraphics[width=\linewidth]{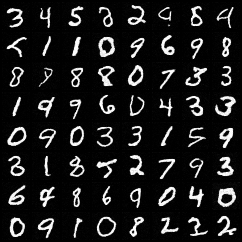}
    \caption{ \scriptsize 
    1HQConv QVU-Net  \\
    Params= 475329 (-1.66\%)\\
    FID= 45.3710\\
    KID= 0.0556 $\pm$ 0.0015\\
    IS= 1.9090 $\pm$ 0.0325}
    \label{fig:mnist_1_20}
  \end{subfigure}
   \hfill
    \hfill
  \begin{subfigure}{0.23\textwidth}
    \includegraphics[width=\linewidth]{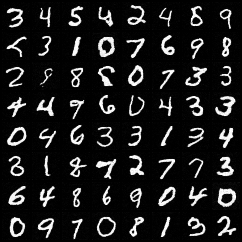}
    \caption{\scriptsize7HQConv QVU-Net  \\
    Params= 440985 (-8.76\%)\\
    FID= 43.0731\\
    KID= 0.0523 $\pm$ 0.0015 \\
    IS= 1.9256 $\pm$ 0.0205 }
    \label{fig:mnist_7_20}
  \end{subfigure}
  \hfill
  \begin{subfigure}{0.23\textwidth}
    \includegraphics[width=\linewidth]{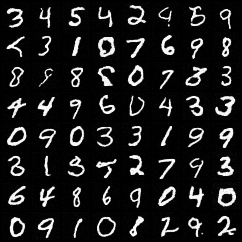}
    \caption{ \scriptsize FullHQConv QVU-Net  \\
    Params= 429993 (-11.03\%)\\
    FID= 45.6503\\
    KID= 0.0559 $\pm$ 0.0015\\
    IS= 1.9053$\pm$ 0.0283}
    \label{fig:mnist_f_20}
  \end{subfigure}
  \caption{MNIST dataset results using the first hybrid architecture. The figure shows in the first column the images generated by the classical network, while in the second column the images generated by the 1HQConv QVU-Net, in the third column by the 7HQConv QVU-Net, and in the last column the images generated by the FullHQConv QVU-Net. The row-wise division considers in the first row the images generated after the networks are trained for just one epoch, the second row after training for ten epochs, and the third row after the complete training of twenty epochs.}
  \label{fig:mnist}
\end{figure}

We then consider the MNIST dataset, whose results are shown in Fig.~\ref{fig:mnist}. In this case, we report the results obtained from the first hybridization, the QVU-Net, with the HQConv ansatz implemented with three layers only, as the results obtained using FQConv or the second hybridization, the QuanvU-Net, that involves the use of quantum in layers other than the vertex are very similar. At the first epoch, hybrid networks have significantly better results than the classical one in terms of all metrics. Similarly to what we observed with the Fashion MNIST dataset, this once again demonstrates the ability of quantum models to perform better than classical models when training epochs are limited. By the tenth epoch, the advantage diminishes somewhat, except for the 7HQConv QVU-Net architecture shown in Fig.~\ref{fig:mnist_7_10} that still shows better values in terms of FID, KID, and IS. At the last epoch, once again the 7HQConv QVU-Net Architecture shows better results as shown in Fig.~\ref{fig:mnist_7_20}, with an obtained FID about two points lower than that of the classical network, along with better IS and KID. In addition to achieving better performance, it is important to note that all hybrid networks have a lower number of parameters than the classical network.

\subsection{Transfer learning}
\begin{figure}[!ht]
  \centering
  \begin{subfigure}{0.23\textwidth}
    \includegraphics[width=\linewidth]{classica_l2_mnist_20.png}
    \caption{ \scriptsize Classical  \\ FID= 45.2762\\
    KID= 0.0558 $\pm$ 0.0014\\
    IS= 1.9152 $\pm$ 0.0355}
    \label{fig:classica}
  \end{subfigure}
   \hfill
  \begin{subfigure}{0.23\textwidth}
    \includegraphics[width=\linewidth]{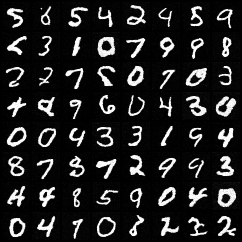}
    \caption{ \scriptsize Transfer learning\\
    FID= 48.9228\\ 
    KID= 0.0600 $\pm$ 0.0014\\
    IS= 1.9094 $\pm$ 0.0241}
    \label{fig:class-class}
  \end{subfigure}
  \hfill
  \begin{subfigure}{0.23\textwidth}
    \includegraphics[width=\linewidth]{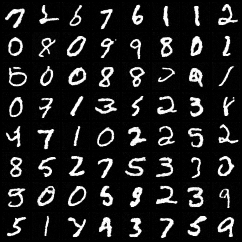}
    \caption{ \scriptsize Transfer learning\\ FID= 41.6646\\ 
    KID= 0.0493 $\pm$ 0.0013\\
    IS= 1.9556 $\pm$ 0.0335}
    \label{fig:class-1fq}
  \end{subfigure}
  \caption{Transfer learning results on the MNIST dataset where generated images are: a) classical architecture; b) transfer learning (19+1) classical-classical; c) transfer learning (19+1) classical-1FQConv QVU-Net.}
  \label{fig:mnist2}
\end{figure}

We can now analyze the results obtained with the transfer learning approach. We compare the metrics of the MNIST dataset images obtained by first training a classical U-Net network for all the 20 epochs and then training a classical network for 19 epochs and transferring the weights except for the vertex to another classical network that is fully retrained for an additional epoch. Finally, we compare the results obtained by training a classical network for 19 epochs and performing transfer learning on the hybrid network 1FQConv QVU-Net, which is retrained for a single epoch.
The results obtained are shown in Fig.~\ref{fig:mnist2}, demonstrating how the transfer learning approach from classical to hybrid works very well.
What is noticeable is that the results obtained in this way are even better than those obtained previously in Fig.~\ref{fig:mnist_7_20}. 

\begin{figure}[!ht]
  \centering
  \begin{subfigure}{0.23\textwidth}
    \includegraphics[width=\linewidth]{classica_l2_20.png}
    \caption{ \scriptsize Classical  \\ FID= 39.4563\\
    KID= 0.0275 $\pm$ 0.0008\\
    IS= 3.8076 $\pm$ 0.0741}
    \label{fig:classica2f}
  \end{subfigure}
   \hfill
  \begin{subfigure}{0.23\textwidth}
    \includegraphics[width=\linewidth]{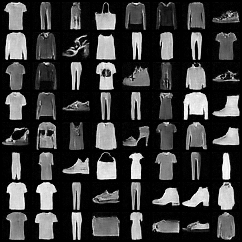}
    \caption{ \scriptsize Transfer Learning\\ FID= 40.6769\\ 
    KID= 0.0273 $\pm$ 0.0008\\
    IS= 4.0764 $\pm$ 0.0945}
    \label{fig:class-class2f}
  \end{subfigure}
  \hfill
  \begin{subfigure}{0.23\textwidth}
    \includegraphics[width=\linewidth]{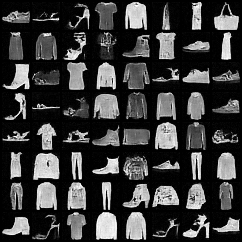}
    \caption{ \scriptsize Transfer Learning\\ FID= 38.6835\\ 
    KID= 0.0261 $\pm$ 0.0007\\
    IS= 4.0527 $\pm$ 0.0890}
    \label{fig:class-1fqf2f}
  \end{subfigure}
   \caption{Transfer learning results on the Fashion MNIST dataset where generated images are: a) classical architecture; b) transfer learning (18+2) classical-classical; c) transfer learning (18+2) classical-1FQConv QVU-Net.}
  \label{fig:fashion2}
\end{figure}

Similarly, the same approach was taken for the Fashion MNIST dataset, with the results reported in Fig.~\ref{fig:fashion2}. Instead of having 19 training epochs on the classical network and just one on the network to which the weights have been transferred, we now have 18 epochs on the classical network and 2 on the network to which the weights have been transferred.
In this case, we observe that the best results are achieved in Fig.~\ref{fig:class-1fqf2f}, even surpassing those obtained in Fig.~\ref{fig:quan_fashion_20}.

Finally, all the numerical results obtained in the previous experiments are summarized in Table~\ref{tab:Fashion_MNIST} and Table~\ref{tab:MNIST} for Fashion MNIST and MNIST datasets, respectively.

\begin{table}[!ht]
  \centering
	  \caption{Numerical results for the images generated from the Fashion MNIST dataset}
   \resizebox{\textwidth}{!}{%
  \begin{tabular}{lcccccc}
    \toprule
     Metrics & Classical & 1HQConv QVU-Net & 7HQConv QVU-Net & FullHQConv QVU-Net & QuanvU-Net & \makecell{Transf.\\Learning}\\
    \midrule
    {Params}& 483321 & 475329& 440985 & 429993 & 474293 & 483321\\
    FID & 39.4563 & 39.9935 & 40.3685& 41.3882& 38.8 & \textbf{38.6835} \\
    KID & 0.0275 $\pm$ 0.0008 & 0.0278 $\pm$ 0.0008 & 0.0281 $\pm$ 0.0009 & 0.0294 $\pm$ 0.0009 & 0.0269 $\pm$ 0.0007& \textbf{0.0261 $\pm$ 0.0007}\\
    IS & 3.9783 $\pm$ 0.0777 &  3.9787 $\pm$ 0.0550 & 3.8158 $\pm$ 0.1360 & 3.8019 $\pm$ 0.0744 & 3.9087 $\pm$ 0.1485 & \textbf{4.0527 $\pm$ 0.0890} \\
    \bottomrule
  \end{tabular}}
  \label{tab:Fashion_MNIST}
\end{table}

\begin{table}[!ht]
  \centering
	 \caption{Numerical results for the images generated from the MNIST dataset}
    \resizebox{\textwidth}{!}{%
  \begin{tabular}{lcccccc}
    \toprule
     Metrics & Classical & 1HQConv QVU-Net & 7HQConv QVU-Net & FullHQConv QVU-Net & QuanvU-Net & \makecell{Transf.\\Learning}\\
    \midrule
    {Params}& 483321 & 475329& 440985 & 429993 & 474293 & 483321\\
    FID & 45.2762 & 45.3710 & 43.0731 & 45.6503 & 44.5144 & \textbf{41.6646} \\
    KID & 0.0558 $\pm$ 0.0014 & 0.0556 $\pm$ 0.0015 & 0.0523 $\pm$ 0.0015 & 0.0559 $\pm$ 0.0015 & 0.0545 $\pm$ 0.0015& \textbf{0.0493 $\pm$ 0.0013}\\
    IS & 1.9152 $\pm$ 0.0355 & 1.9090 $\pm$ 0.0325 & 1.9256 $\pm$ 0.0205 & 1.9053 $\pm$ 0.0283 & 1.9284 $\pm$ 0.0425 & \textbf{1.9556$\pm$ 0.0335} \\
    \bottomrule
  \end{tabular}}
   \label{tab:MNIST}
\end{table}


\section{Conclusions}
\label{sec: conclusions}
The use of quantum computing in generative machine learning models can bring numerous advantages, both in terms of performance and in terms of reducing the parameters to be trained. In this paper, we proposed an efficient integration of quantum computing in diffusion models, presenting for the first time two possible hybrid U-Nets. The first, the QVU-Net, involves replacing the convolutional layers that form the ResNet with variational circuits only at the vertex, while the second, the QuanvU-Net, involves the replacement in the second block of the encoder part as well, leveraging in this case an approach inspired by quanvolutional.
We also attempted to exploit an approach inspired by transfer learning to reduce the overall training times compared to what we would have had with the complete training of a hybrid architecture.

The results obtained confirm that there is a real advantage in using quantum in extremely complex networks, such as the U-Net of DMs. The approach of hybridizing the U-Net confirms that integrating variational circuits into a classical network can yield certain benefits. Through numerous tests, we proved that quantum allows for further enhancement of network performance. For MNIST, the use of the 7HQConv QVU-Net yielded the best performance. Furthermore, not only at the twentieth epoch did the use of quantum lead to better performance, but in general all hybrid networks show significantly more positive metric values compared to the classical network they are consistently compared against from the first epoch.

On Fashion MNIST, the first possible hybridization of the U-Net, the QVU-Net,s which involves only the hybridized vertex, fails to yield better results than the classical one, despite having a significantly lower number of parameters than the classical network. However, the quantum advantage in this case lies in a faster learning rate, as when we analyze the results at the tenth epoch, all hybrid networks still demonstrate metric values much more advantageous than those observed from the classical network. It is with the second implementation, the QuanvU-Net, which also involves the use of quanvolutional in more expressive layers, that better results are achieved. This demonstrates that having the introduction of quantum in areas dedicated to feature extraction is more effective than introducing it only at the vertex, which primarily operates on processing features already extracted earlier.

Finally, the idea behind using transfer learning between a classical network and a hybrid network is driven by the desire to keep simulation times limited while still achieving better performance than the classical network. This is observed both in the case of MNIST and Fashion MNIST, where we indeed obtain the best results. It is essential to note that the main goal of this paper is to demonstrate that quantum outperforms or performs equally to a classical network with more parameters. It is worth emphasizing that all hybrid networks have significantly fewer parameters.

The possible future developments of this work involve expanding hybridization to other parts of the U-Net, replacing convolutional layers with variational circuits even in regions where the image is less downscaled. This aims to achieve further reduction in the number of trainable parameters, in addition to the potential for improved performance. Furthermore, the goal is to extend the testing to more complex datasets beyond MNIST and Fashion MNIST.

\section*{Acknowledgments}
The contribution in this work of M. Panella, A. Ceschini and F. De Falco was in part supported by the ``NATIONAL CENTRE FOR HPC, BIG DATA AND QUANTUM COMPUTING'' (CN1, Spoke 10) within the Italian ``Piano Nazionale di Ripresa e Resilienza (PNRR)'', Mission 4 Component 2 Investment 1.4 funded by the European Union - NextGenerationEU - CN00000013 - CUP B83C22002940006.

\end{document}